# Simple and accurate expressions for radial distribution functions of hard disk and hard sphere fluids

**Hongqin Liu (刘洪勤)***

Integrated High Performance Computing, Shared Services Canada, Montreal, Canada

Analytical expressions for radial distribution function (RDF) are of critical importance for various applications, such as development of the perturbation theories for equilibrium properties. Theoretically, RDF expressions for odd-dimensional fluids can be obtained by solving the Percus-Yevick integral equations. But for even-dimensional cases, such as the hard disk (2D) fluid, analytical expressions are infeasible. The only 2D RDF is a heuristic expression proposed by Yuste et al. (J. Chem. Phys. 99, 2020,1993), which approximates the 2D RDF with an interpolation of the RDFs for hard-rod (1D) and hard-sphere (3D) fluids and provides acceptable estimations for an intermediate and low density range. In this work, we employ a simple and empirical expression for the 2D RDF and the 3D RDF based on the approach proposed by Trokhymchuk et al. for the 3D RDF (J. Chem. Phys., 123, 024501, 2005). The parameters are determined in such a way that the final RDF expressions are thermodynamically consistent, namely the pressure constraint and the isothermal compressibility constraint are both satisfied. The new RDFs for the 2D and 3D hard spheres are highly accurate for the entire density range up to the first-order phase transition points. The predictions of the first coordination numbers are consistent with simulation results for 3D fluid. Finally, by using the 2D RDF with a primitive second-order perturbation theory, the pressure-volume-temperature relation and vapor-liquid equilibrium are calculated for the 2D Lennard-Jones fluid. Comparisons with the simulation data show promising results.

* Emails: hongqin.liu@ssc-spc,gc.ca; hqliu2000@gmail.com.

## I. Introduction

Radial distribution function (RDF) plays a key role in describing structural and equilibrium properties of fluids. Analytical expression of a radial distribution function is of critical importance for various applications, such as in the perturbation theories, where the perturbation terms are expressed in terms of the RDF [1,2,3]. For the hard sphere (3D) fluid, analytical expressions can be obtained by solving the Orstein-Zernike equation with the Percus-Yevick (PY) approximation (the PY integral equation). The PY equation is exactly solvable in odd dimensions, such as 1D (hard rod) and 3D fluids. Several theoretical expressions for the 3D RDF have been proposed. For example, Yuste and Santos derived an expression for the 3D RDF using the Rational Function Approximation (RFA) method [4]. Some other expressions are also reported [5,6,7,8].

Almost all theoretical expressions (i.e. 3D RDF) are mathematically lengthy and inconvenient for practical applications. For instance, in perturbation theories, integrations of an RDF w.r.t. the distance variable (radial coordinate) are required [1,2,3]; and in determining the effective hard sphere diameter for the Lennard-Jones fluid, derivative of the RDF w.r.t. diameter is demanded with Lado's perturbation theory [9]. Apparently, with a lengthy RDF, the above applications will become cumbersome and simple expressions are always desirable. For this reason, Trokhymchuk et al. [10] proposed a simple expression (TNJH) with the parameters obtained by imposing some physical constraints. López de Haro et al. [11] discussed the TNJH expression and pointed out that it's accuracy is lower than that of the RFA expression [4].

For the 2D (hard disk) fluid, since theoretical expression is infeasible [12], the only analytical expression is an approximation proposed by Yuste and Santos 30 years ago [12]. This expression employs an interpolation of the strict expression of the 1D RDF and the RFA expression for the 3D RDF [4]. As shown by ref.[12] (FIG.4 and FIG. 5 therein), at high densities ( >0.7) the prediction accuracy becomes poor. In addition, the continuity of the interpolation is only at the zeroth order, namely the derivative of the 2D RDF w.r.t the radial coordinate is discontinuous. Considering the significance of the 2D fluid, a new analytical and accurate 2D RDF is much needed for reliably describing both the structural and thermodynamic properties over a wide density range.

In this work, we use a simplified form of the TNJH RDF [10] for both 2D and 3D fluids. Fewer parameters are involved and the physical constraints used by the TNJH RDF [10] are imposed to obtain the values of the parameters. The simple form allows straightforward analytical integrations w.r.t. the distance variable, or derivatives w.r.t. diameter. Most importantly, it turns out that the final expressions are highly accurate for both 2D and 3D RDFs over wide density ranges up to the first-order phase transition points. For demonstrating the applications of the 2D RDF, the two dimensional Lennard-Jones (2DLJ) fluid is discussed with a simple perturbation theory in section V.

## II. General background

Here we adopt the strategy proposed by Trokhymchuk et al.[10]. The RDF is divided into a depletion branch and a structural branch:

$$g(r) = \begin{cases} 0, & r < \sigma \\ g^{dep}(r), \sigma \leq r < r_m \\ g^{str}(r), r_m \leq r < \infty \end{cases} \quad (1)$$

where $r$ is the radial coordinate; $\sigma$, the diameter of the hard sphere; $r_m$, a point at which the two branches are united and is taken as a parameter to be determined. In the original TNJH model, the expression for $g^{dep}(r)$ was from a theoretical expression [10]:

$$g^{dep}(r) = \frac{A}{r}e^{\mu(r-\sigma)} + \frac{B}{r}\cos[\beta(r-\sigma)+\gamma]e^{\alpha(r-\sigma)} \quad (2)$$

Apparently, 6 parameters ($A, B, \beta, \mu, \gamma, \alpha$) are involved in the equation. It turns out the cosine term is unnecessary since this term deals only with the first peak, and hence we propose the following expressions for the depletion and structural branches, respectively:

$$g^{dep}(r) = \frac{A}{r}e^{\mu(r-\sigma)} + \frac{B}{r}e^{\alpha(r-\sigma)} \quad (3)$$

$$g^{str}(r) = 1 + \frac{C}{r}\cos(\omega r + \delta)e^{-\kappa r} \quad (4)$$

The expression for $g^{str}(r)$, Eq.(4), is the same as that in the TNJH RDF [10]. Equations (1), (3) and (4) contains 8 parameters (4 from the structural term, $C, \omega, \delta, \kappa$) in total, 2 fewer than that in the original TNJH expression, Eq.(2). Before proceeding, we first provide some background for the most important quantities to be used in the following sections. The radial distribution function at contact, $g(\sigma) = g(r = \sigma)$, is related to the compressibility $Z = P/(\rho k_B T)$ by

$$g(\sigma) = \frac{Z-1}{2^{D-1}\eta} \quad (5)$$

where $P$ is pressure; $\rho$, number density; $k_B$, the Boltzmann constant; $T$, temperature. The generic packing fraction is defined as:

$$\eta = \frac{\pi^{\frac{D}{2}}}{\Gamma(1+D/2)} \rho \left(\frac{\sigma}{2}\right)^D \quad (6)$$

which gives $\eta = \pi\rho\sigma^2/4$ for hard disk (D=2), $\eta = \pi\rho\sigma^3/6$ for hard sphere (D=3). The reduced (dimensionless) isothermal compressibility is defined as:

$$\chi_T = k_B T \left(\frac{\partial \rho}{\partial P}\right)_T = \frac{1}{Z + \eta Z'} \quad (7)$$

where $Z' = \partial Z/\partial \eta$. Equation (5) and (7) tell that given an analytical equation of state (EoS), $g(\sigma)$ and $\chi_T$ can be obtained. On the other hand, as an RDF is given, we can write the pressure as (known as the virial equation):

$$\frac{P}{\rho k_B T} = 1 + 2^{D-1}\eta g(\sigma) \quad (8)$$

which is in fact Eq.(5). The compressibility equation reads

$$k_B T\left(\frac{\partial \rho}{\partial P}\right)_T = 1 + 2^{D-1}\pi\rho \int_0^\infty [g(r)-1] r^{D-1} dr \quad (9)$$

The so-called thermodynamic consistency is defined such that an RDF, $g(r)$, satisfies both Eq.(8) and (9). With the method discussed in this work, we assume a priori knowledge of an accurate EoS for determining $g(\sigma)$ and $\chi_T$. We are now ready for presenting various constraints [10] for the parameter-estimations. The first one is from Eq.(8) (or Eq.(5)), namely,

$$g^{dep}(\sigma) = g(\sigma) \quad (10)$$

The second is from Eq.(9) and (7):

$$\int_0^\infty [g(r)-1] r^{D-1} dr = \frac{\chi_T - 1}{2^{D-1}\pi\rho} \quad (11)$$

As mentioned, in Eq.(10) and Eq.(11), $g(\sigma)$ and $\chi_T$ will be calculated from an EoS, and the later is usually derived from fitting computer simulation data for the compressibility and virial coefficients [13,14]. Therefore, these two quantities ($g(\sigma)$ and $\chi_T$ from an EoS) can be seen as "experimental" results [10]. Now we define a kissing point $r = r_m$, where:

$$g^{dep}(r_m) = g_m \quad (12a)$$
$$g^{str}(r_m) = g_m \quad (12b)$$

$r_m$ and $g_m$ are treated as parameters. For a smooth transition from $g^{dep}(r)$ to $g^{str}(r)$ at $r_m$, we force it to be the first minimum point of the RDF, namely:

$$\left[\frac{d}{dr} g^{dep}(r)\right]_{r=r_m} = 0 \quad (13a)$$

$$\left[\frac{d}{dr} g^{str}(r)\right]_{r=r_m} = 0 \quad (13b)$$

Therefore, we have in total 6 constraints with 8 + 2 ($r_m$ and $g_m$) unknowns. Since we will fit the model globally at multiple densities, the mismatch of the number of unknowns with that of the equations does not cause any issue. Figure 1 shows how the depletion branch is united with the structural branch at $r_m$ while satisfying constraints, Eq.(12) and Eq.(13). From the figure we can see that removal of the cosine function in the depletion branch, Eq.(2), is mathematically justified since no periodic behavior is present for the first peak. All other peaks are taken care of by the structural branch.

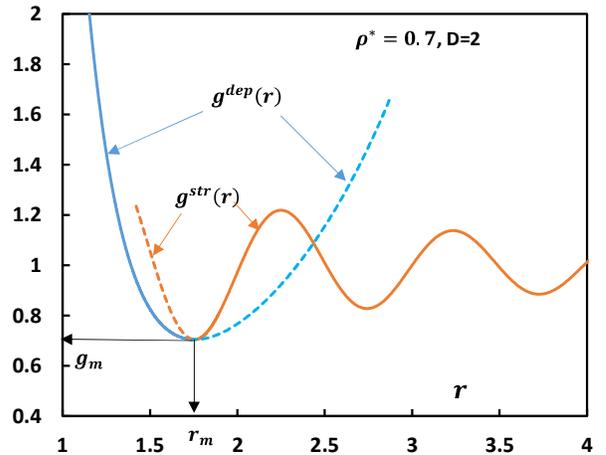

**Figure 1**. Illustration of the unification of the depletion and structural branches at $r = r_m$.

To minimize the number of unknowns in a global optimization process, from Eq.(10)-(13), we can explicitly express some parameters in terms of density, $r_m$ and $g_m$, etc. First of all, at $r = \sigma$ and $r = r_m$, respectively, we have:

$$\frac{B}{\sigma} = \frac{r^* g_m - g(\sigma) e^{\mu\sigma(r^*-1)}}{e^{\alpha\sigma(r^*-1)} - e^{\mu\sigma(r^*-1)}} \quad (14)$$

$$\frac{A}{\sigma} = g(\sigma) - \frac{B}{\sigma} \quad (15)$$

where $r^* = r_m/\sigma$. In all the equations presented here, by following the conventions used in Ref.[10], $\mu\sigma$, $\alpha\sigma$, $A/\sigma$ etc. should be regarded as a single parameter. Meanwhile, from Eq.(12a) and Eq.(13a), after some straightforward algebra, we have

$$\frac{A}{\sigma} \mu\sigma e^{\mu\sigma(r^*-1)} + \frac{B}{\sigma} \alpha\sigma e^{\alpha\sigma(r^*-1)} = g_m \quad (16)$$

Replacing $A/\sigma$ and $B/\sigma$ with Eq.(14) and (15), respectively, we have an equation that relates $\mu\sigma$ to $\alpha\sigma$:

$$\frac{1}{e^E}(x+E)e^{(x+E)} + F = 0 \quad (17)$$

where $x \equiv \mu(r^* - 1)$, and

$$E \equiv \frac{[g_m - \alpha e^{\alpha\sigma(r^*-1)} g(\sigma)](r^* - 1)}{e^{\alpha\sigma(r^*-1)} g(\sigma) - r^* g_m} \quad (18a)$$

$$F \equiv \frac{g_m(\alpha\sigma r^* - 1)(r^* - 1) e^{\alpha\sigma(r^*-1)}}{e^{\alpha\sigma(r^*-1)} g(\sigma) - r^* g_m} \quad (18b)$$

Let $y = x + E$, we have:

$$y e^y = -F e^E \quad (19)$$

The right side of Eq.(19) is a function of $\alpha\sigma$, $g_m$ and $r^*$, and it can be solved with the Lambert W function, and we finally have:

$$\mu\sigma = \frac{W(-Fe^E) - E}{r^* - 1} \quad (20)$$

Eq.(20) explicitly express $\mu\sigma$ in terms of $\alpha\sigma$. Therefore, we have one unknown less to be determined by the global optimization. Apparently, if the original TNJH expression, Eq.(2), is adopted, such a dependence, Eq.(20), does not hold. We can also determine some relations for the structural branch. From Eq.(12b) and Eq.(13b), it is straightforward to obtain the following relations [10]:

$$\delta = -\omega\sigma r^* - arctan\left(\frac{\kappa\sigma r^* + 1}{\omega\sigma r^*}\right) \quad (21)$$

$$\frac{C}{\sigma} = \frac{r^*[g_m - 1]e^{\kappa\sigma^*}}{cos(\omega\sigma r^* + \delta)} \quad (22)$$

In summary, by using Eq.(14), Eq.(15), Eq.(20), Eq.(21) and Eq.(22) to eliminate 5 unknowns, 4 parameters, $\alpha\sigma$, $\kappa\sigma$, $r^*$ and $g_m$ are left to be determined. We still have the isothermal compressibility constraint, Eq.(11), available. This will be discussed for 2D and 3D cases, respectively, since the detailed expressions are D-dependent. By using the simulation data at multiple densities, the 4 unknowns are evaluated by a global optimization. The objective function to be minimized is composed of the isothermal compressibility constraint, plus a distance between the calculated RDF and the simulation data, $\sum \|g^{cal}(r) - g^{sim}(r)\|$. Appropriate weighting between the two is applied for obtaining best results. The final results are presented in succeeding sections.

## III. RDF for the hard disk fluid

As mentioned, we still have the isothermal compressibility constraint to be finalized. For the 2D RDF, Eq.(9) becomes

$$\chi_T = 1 - \pi\rho^* r^{*2} + 2\pi\rho^*\left(\frac{J_1}{\sigma^2} + \frac{J_2}{\sigma^2}\right) \quad (23)$$

where $\rho^* = \rho\sigma^2$. The integrations in Eq.(9) can be easily carried out analytically:

$$J_1 = \int_\sigma^{r_m} g^{dep}(r) r \, dr = \frac{A}{\mu}\left[e^{\mu(r_m-\sigma)} - 1\right] + \frac{B}{\alpha}\left[e^{\alpha(r_m-\sigma)} - 1\right] \quad (24)$$

$$J_2 = \int_{r_m}^\infty [g^{str}(r) - 1] r \, dr =$$
$$-\frac{Ce^{-\kappa r_m}}{\omega^2 + \kappa^2}[\kappa cos(\omega r_m + \delta) - \omega sin(\omega r_m + \delta)] \quad (25)$$

In addition, we can also calculate the first coordination number:

$$n_1 = 2\pi\rho^* \frac{J_1}{\sigma^2} \quad (26)$$

For calculation of the compressibility, $Z$, and hence the RDF at contact, $g(\sigma)$, and the isothermal compressibility, $\chi_T$, we need an accurate EoS for the 2D fluids. Recently, the present author proposed an EoS for the entire density range, from zero density to the hexatic phase [13,14]:

$$Z_{lh} = Z_v + Z_p = \frac{1 + \frac{1}{8}\eta^2 + \frac{1}{18}\eta^3 - \frac{4}{21}\eta^4}{(1-\eta)^2} + \frac{b_1\eta^{m_1} + b_2\eta^{m_2}}{1 - c\eta} \quad (27)$$

Where, $Z_{lh}$ refers to the compressibility for the liquid and hexatic phases, $Z_v$ is a Carnahan-Starling type EoS derived from the virial coefficients [13], $Z_p$ is the contribution from a term with a pole at $\eta = 1/c$. It is demonstrated in Ref.[14] that Eq.(27) can not only accurately reproduce the simulation data for pressure, it can also accurately produce the liquid-hexatic phase transition. As a matter of fact, it works well until the hexatic-solid transition point, hence functions as the liquid-hexatic branch of a global EoS for the 2D fluid and solid [14]. Figure A1 (Appendix A) depicts the entire density spectrum for the entire range from the gas to solid phases [14] for reference.

In a density range not close to the liquid-hexatic transition, $\rho^* < 0.8$, $Z_v$ is accurate enough, but for higher density range until the hexatic phase, $\rho^* = 0.9167$, the contribution from $Z_p$ has to be considered [14]. The values of the EoS constants in Eq.(27) are listed in Table A1 (Appendix A). The expression for the isothermal compressibility can be easily derived from the EoS. As shown in Ref.[11], the isothermal compressibility can be reliably predicted by an accurate EoS and the values for $\chi_T$ obtained can be treated as the computer "experimental" data [10].

Plenty of simulation data for the 2D RDF can be found in the literature [15,16,17,18,19,20], that cover wide density ranges from $\rho^* = 0.46$ to $\rho^* = 0.9$. Most importantly, as shown below, these data are fairly consistent with each other. These facts make the parameter-estimation reliable. In this work, we are particularly interested in the high density range. For this reason, we employed the data from Ref.[15]-Ref.[19] for parameter-evaluations, and data from Ref.[20] (high density-range) for testing only. The parameters are finally expressed as polynomial functions of the packing fraction:

$$\omega\sigma = 4.07688 + 3.53995\eta \quad (28)$$

$$\kappa\sigma = 1.19878 - 1.54059\eta - 0.567808\eta^2 \quad (29)$$

$$\alpha\sigma = -1.69377 + 13.50042\eta - 29.8772\eta^2 \quad (30)$$

$$r^* = 2.14098 - 1.03267\eta + 0.320164\eta^2 + 0.0102421\eta^3 \quad (31)$$

$$g_m = 1.97652 - 6.34821\eta + 11.53962\eta^2 \\ -5.96960\eta^3 - 3.92125\eta^4 \quad (32)$$

Another parameter, $\mu\sigma$, can be calculated from the Lambert W function, Eq.(20), given Eq.(29)-Eq.(32). For user's convenience, here we fitted the results from Eq.(20) with the following polynomial function in the density range $\rho^* = 0.4 - 0.9$:

$$\mu\sigma = -116.2837 + 1218.065\eta - 4784.826\eta^2 \\ +9229.068\eta^3 - 8862.922\eta^4 + 3422.542\eta^5 \quad (33)$$

In summary, Eq.(1), Eq.(3), Eq.(4) together with Eq.(14), Eq.(15), Eq.(21), Eq.(22), and Eq.(28)-Eq.(33) comprise the RDF for the hard disk fluid for the density range $\rho^* = 0.4 - 0.9$. The EoS that provides $g(\sigma)$ and $\chi_T$ is given by Eq.(27). This RDF will reproduce precisely the same compressibility (or pressure) as given by the EoS, which is guaranteed by Eq.(14) and Eq.(15). Figure 2 illustrates the isothermal compressibility from the RDF compared with the EoS results. The Figure reveals that as density $\rho^* > 0.85$, the isothermal compressibility from the RDF quickly falls to zero. In other words, as the density approach the super-cooled liquid region, the general correlations, Eq.(28)-Eq.(32), fail to represent the density dependence. However, the thermodynamic consistency constraint can be reserved without using the correlations. The RDF produced with above expressions is still valuable since the pressure constraint is strictly met. In real world applications, isothermal compressibility should be always calculated from an EoS, not from an RDF.

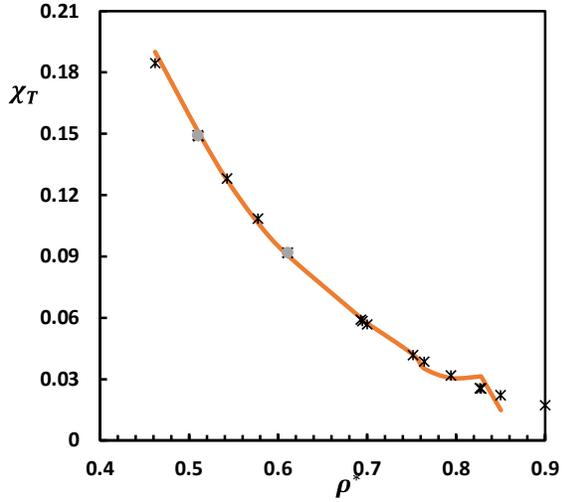

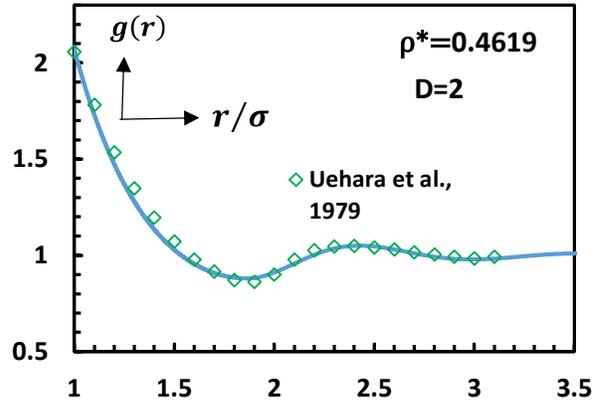

**Figure 2**. Comparison of the reduced isothermal compressibility from the RDF (solid line) with the EoS calculations (points).

Figure 3 depicts the first coordination number computed from Eq.(26). As shown by the figure, as density $\rho^* \to 0.85$, $n_1 \to 5.8$. This is a reasonable prediction by noticing that the perfect packing number is 6.

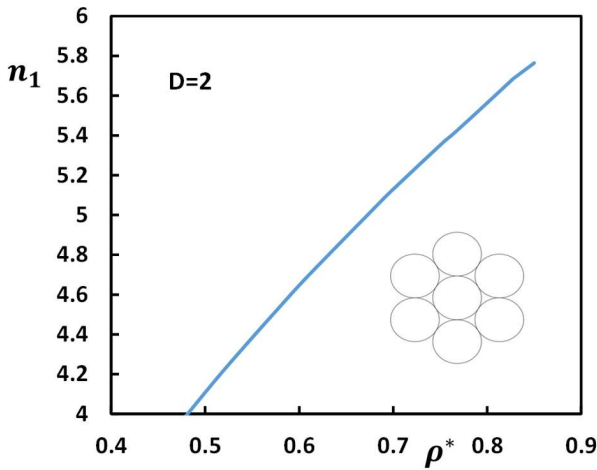

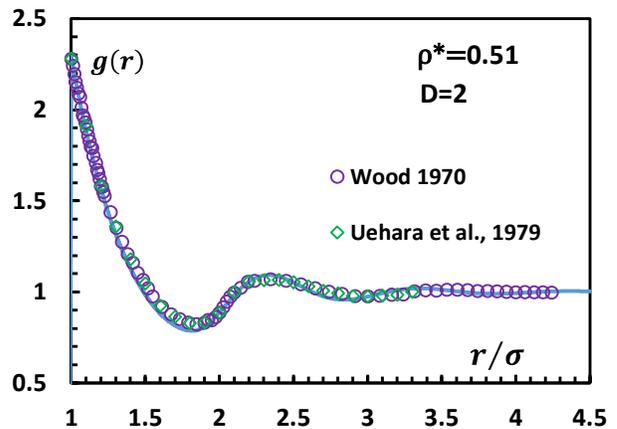

**Figure 3**. The first coordination number calculated by Eq.(26). The inner set shows the perfect packing.

For demonstrating the accuracies of the 2D RDF, Figure 4 depicts the results for some selected densities from the lowest ($\rho^* = 0.4619$) to a high density, $\rho^* = 0.85$. As shown by the figures, the agreements between the calculated RDF values and simulation results are generally excellent.

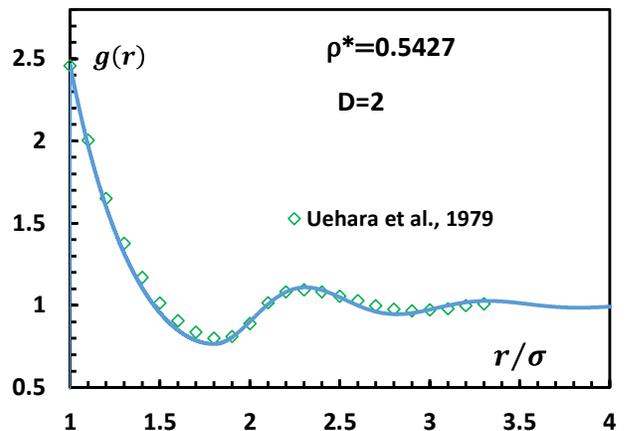

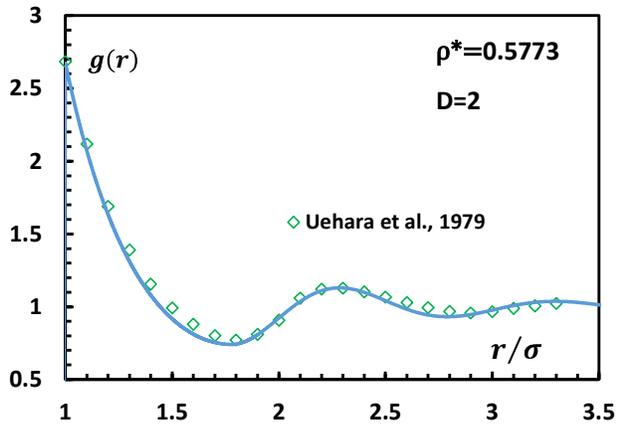
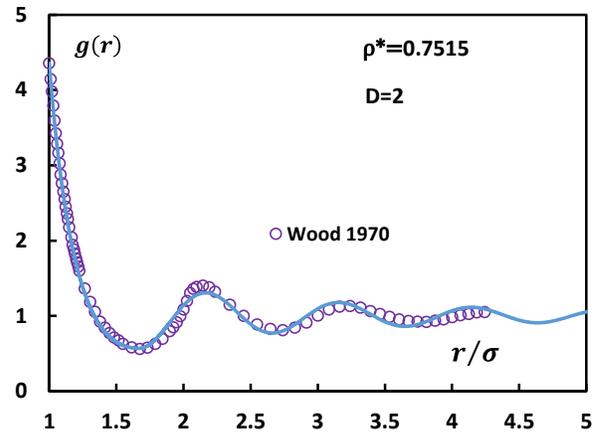
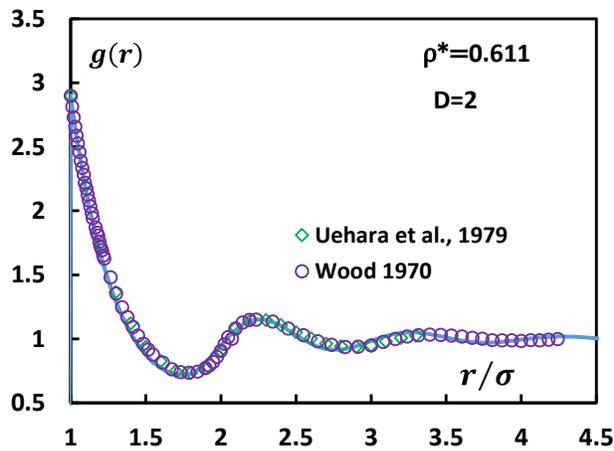
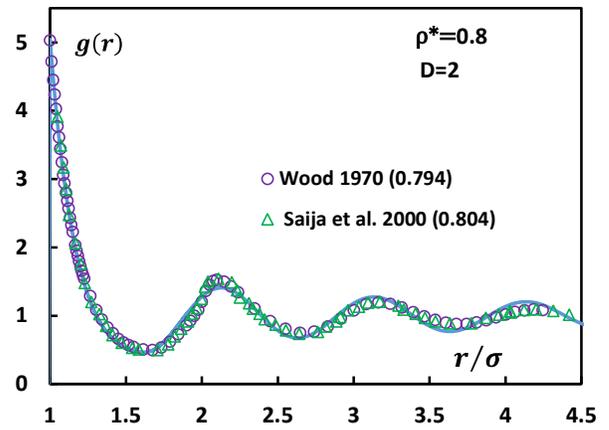
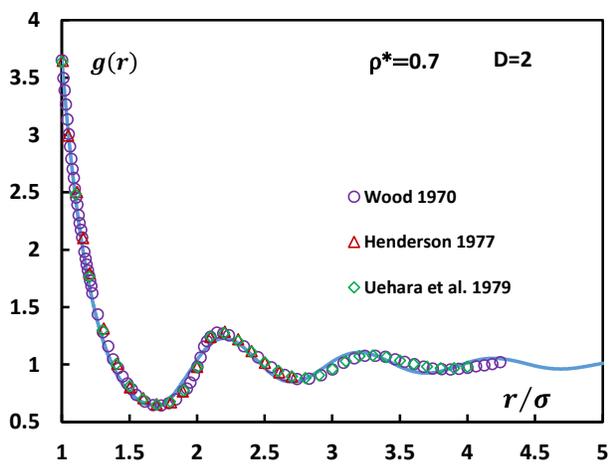
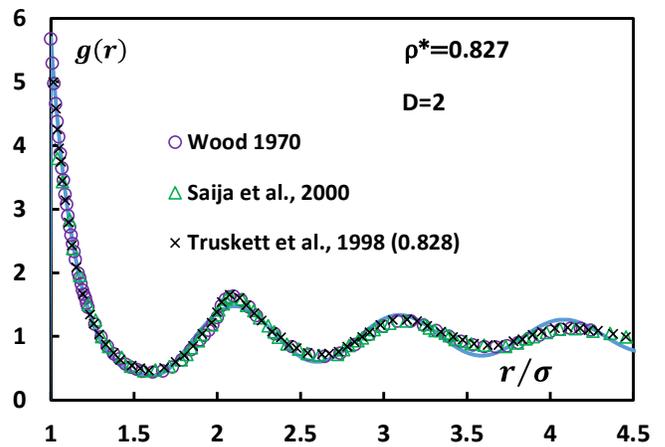

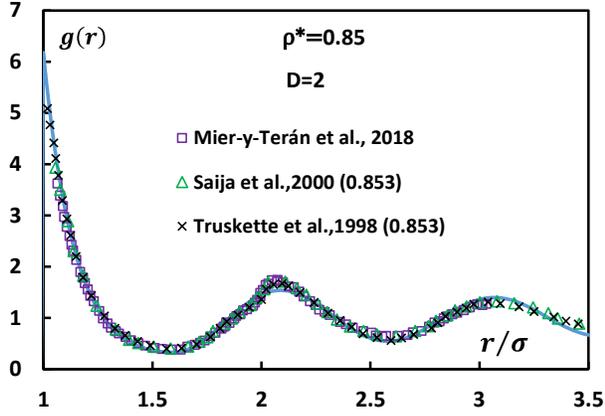

**Figure 4**. Comparisons of the 2D RDF expression (solid lines) with simulation data (points). Data sources are: $\rho^* = 0.4619$ ref.[17], $\rho^* = 0.51$ ref.[15,17], $\rho^* = 0.5773$ ref.[17], $\rho^* = 0.611$ ref.[15,17], $\rho^* = 0.7$ ref.[15,16,17], $\rho^* = 0.7515$ ref.[15], $\rho^* = 0.8$ ref.[15,18], $\rho^* = 0.827$ ref.[15,18,20], $\rho^* = 0.85$ ref.[18,19,20].

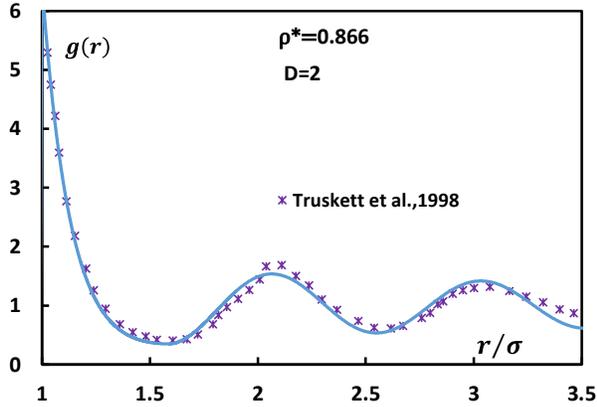

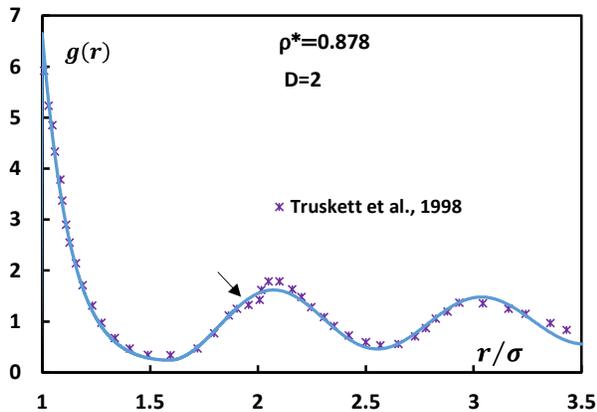

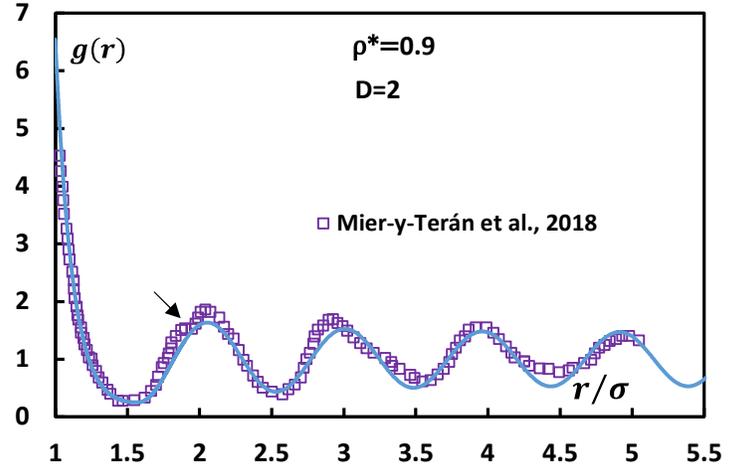

**Figure 5**. The 2D RDF at high densities close to the phase transition. Solid lines are from the analytical expression. Data sources (points) are: $\rho^* = 0.866$ ref.[20], $\rho^* = 0.878$ ref.[20] and $\rho^* = 0.9$ ref.[19]. The arrows indicate shoulders at the second peak.

Finally, Figure 5 illustrates the calculation results for a high density range up to $\rho^* = 0.9$. At density, $\rho^* = 0.878$, a shoulder appears at the second peak, which is a precursor of a super-cooled liquid phase or a new phase (the hexatic phase) [20]. At $\rho^* = 0.9$, the precursor becomes bold with a shoulder starting at $r/\sigma \approx 1.9$. Although the 2D RDF expression can not reproduce any shoulder, the feature of a long-range order is well captured nevertheless. As mentioned, in Figure 4 and Figure 5, data from Truskett et al. (1998) [20] were not included in the parameter-estimations and used for testing purpose only.

As reported by Ref.[14] (refer to Figure A1), EoS calculations reveal that the super-cooled liquid phase appears at $\rho^* = 0.89$, and the system enters a two-phase (liquid and hexatic) region at $\rho^* = 0.9$, and finally at $\rho^* \approx 0.91$, a pure hexatic phase appears. The structural property, RDF, shows consistent results. However, the phase transition density indicated by the RDF (structural property) is slightly lower than that predicted by an EoS (thermodynamic property).

### IV. RDF for the hard sphere fluid

Now we apply the simplified RDF expression, Eq.(1), Eq.(3) and (4) to the hard sphere (3D) fluid. As mentioned, the new depletion branch, Eq.(3), has two-parameter fewer than that of the original TNJH RDF [10], which will considerably simplifies the final model.

For the 3D fluid, the isothermal compressibility is given by

$$\chi_T = 4\pi\rho \int_0^\infty [g(r)-1] r^2 dr = 1 - \frac{4\pi}{3}\rho\sigma^3 r^* + 4\pi\rho J_3 + 4\pi\rho J_4 \quad (34)$$

where the integrations can also be carried out analytically:

$$J_3 = A\frac{1}{\mu^2}(\mu r_m - 1)\exp(\mu(r_m - 1)) - A\frac{1}{\mu^2}(\mu - 1) + B\frac{1}{\alpha^2}(\alpha r_m - 1)\exp(\alpha(r_m - 1)) - B\frac{1}{\alpha^2}(\alpha - 1) \quad (35)$$

$$J_4 = -\frac{Ce^{-\kappa r_m}}{(\omega^2 + \kappa^2)^2} \left\{ \begin{array}{l} \omega(\omega^2 r_m + \kappa^2 r_m + 2\kappa)\sin(\omega r_m + \delta) \\ -[\kappa^2(\kappa r_m + 1) + \omega^2(\kappa r_m + 1)]\cos(\omega r_m + \delta) \end{array} \right\} \quad (36)$$

The first coordination number is given by:

$$n_1 = 4\pi\rho \int_\sigma^{r_m} g^{dep}(r) r^2 dr = 4\pi\rho J_3 \quad (37)$$

For calculations of the RDF at contact (the pressure path) and $\chi_T$ (the reduced isothermal compressibility), a highly accurate EoS is required. Here we use the new Carnahan-Starling type EoS developed by current author based on an accurate correlation of the virial coefficient (up to 12$^{th}$) [13]:

$$Z = \frac{1 + \eta + \eta^2 - \frac{8}{13}\eta^3 - \eta^4 + \frac{1}{2}\eta^5}{(1-\eta)^3} \quad (38)$$

The reduced isothermal compressibility is calculated with the following equation:

$$\chi_T = \frac{(1-\eta)^4}{1 + 4\eta + 4\eta^2 - \frac{32}{13}\eta^3 - \frac{57}{13}\eta^4 + 5\eta^5 - \frac{3}{2}\eta^6} \quad (39)$$

The simulation data for 3D RDF are from Ref.[10] for density range $\rho^* = 0.3 - 0.9$, and Ref.[20,21] ($\rho^* = 0.943, 0.94$). As demonstrated, the high accuracy of the EoS guarantees the reliable prediction of the isothermal compressibility [13]. By using the same approach as discussed for the 2D RDF, the parameters of the 3D RDF are obtained from a global optimization and eventually fitted with the following polynomial functions:

$$\omega\sigma = 6.56745 + 0.721795\eta \quad (40)$$

$$\kappa\sigma = 1.75879 - 3{,}23977\eta + 0.449588\eta^2 - 0.148890\eta^3 \quad (41)$$

$$\alpha\sigma = 0.595157 - 2.469402\eta - 31.99603\eta^2 - 0.897749\eta^3 \quad (42)$$

$$r^* = 1.960492 - 0.240131\eta - 3.26102\eta^2 + 4.08014\eta^3 \quad (43)$$

$$g_m = 1.072954 - 1.13263\eta + 3.05087\eta^2 - 5.94853\eta^3 \quad (44)$$

And the following function is for replacing the W function, Eq.(20), for future applications:

$$\mu\sigma = 0.58236 + 8.43296\eta - 34.52268\eta^2 + 58.63440\eta^3 - 35.13525\eta^4 \quad (45)$$

Compared with the original TNJH 3D RDF [10], the new 3D RDF is significantly simplified. Figure 6 depicts the comparison of the isothermal compressibility values from the RDF (solid line) with the "experimental" data (EoS, points) from Eq.(39). For the pressure (or the RDF at contact), the RDF precisely reproduce the same values as the EoS, due to Eq.(14) and Eq.(15). Therefore, the new RDF is thermodynamically consistent over the entire density range for the hard sphere fluid.

Figure 7 compares the predicted first coordination number, Eq.(37), (solid line) with the simulation data (points) from Ref.[10]. The results show an excellent agreement, and the curve correctly predicts $n_1 \to 13$ at the highest density.

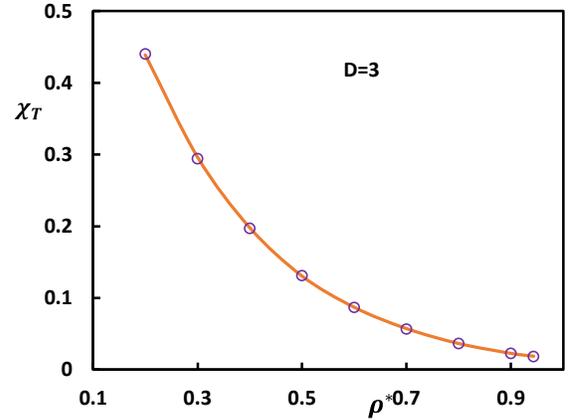

**Figure 6**. Comparison of the reduced isothermal compressibility from EoS, Eq.(7) and Eq.(39) (points), with those from the RDF (solid line).

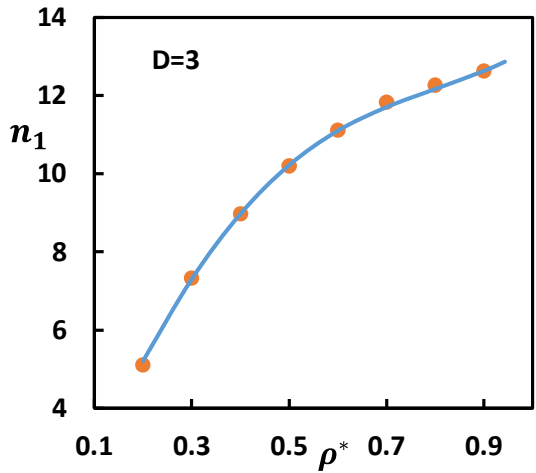

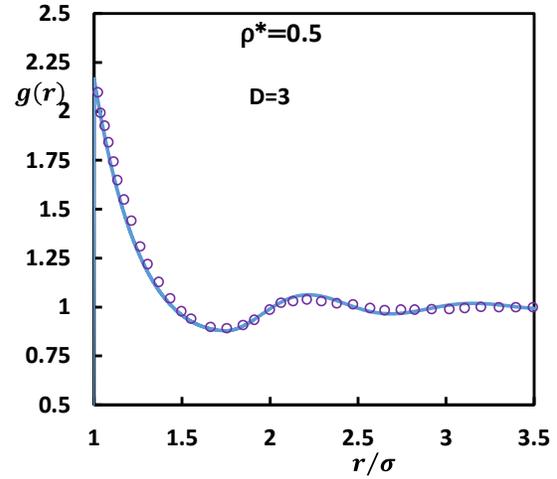

**Figure 7**. The first coordination number calculated from Eq.(37) (line) compared with simulation data (points) [10].

Figure 8 depicts comparisons of calculated RDFs with the simulation data [10,20], at selected densities up to the freezing point, $\rho^* = 0.943$ [20]. The results from the original TNJH RDF [10] are also shown for comparisons at several high densities. With two-parameters less and a simpler form, the new RDF behaves better than the original TNJH expression at high densities. Several factors may contribute to the improvement: (1) a more accurate EoS, Eq.(38); (2) the Lambert W function, Eq.(20), helps reducing the number of unknowns in the global optimization; (3) an appropriately defined objective function with weighted balance between the isothermal compressibility constraint and a RDF distance function.

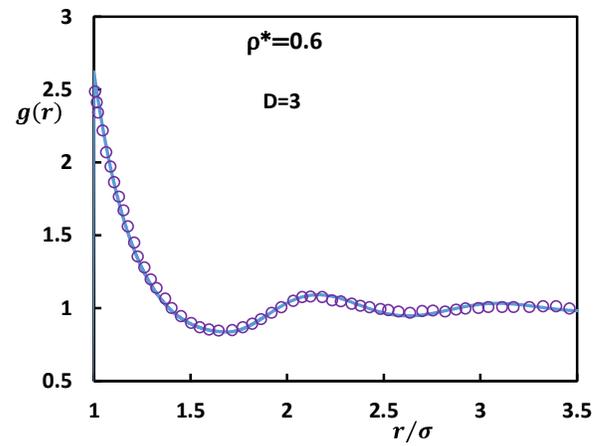

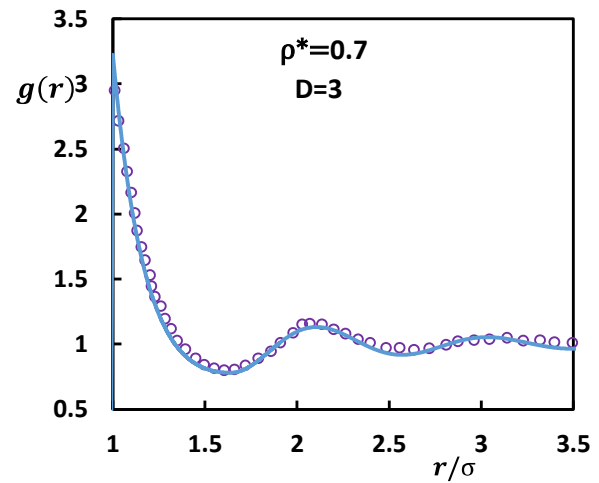

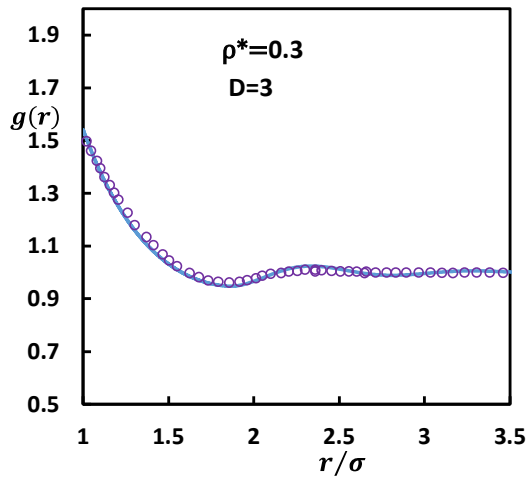

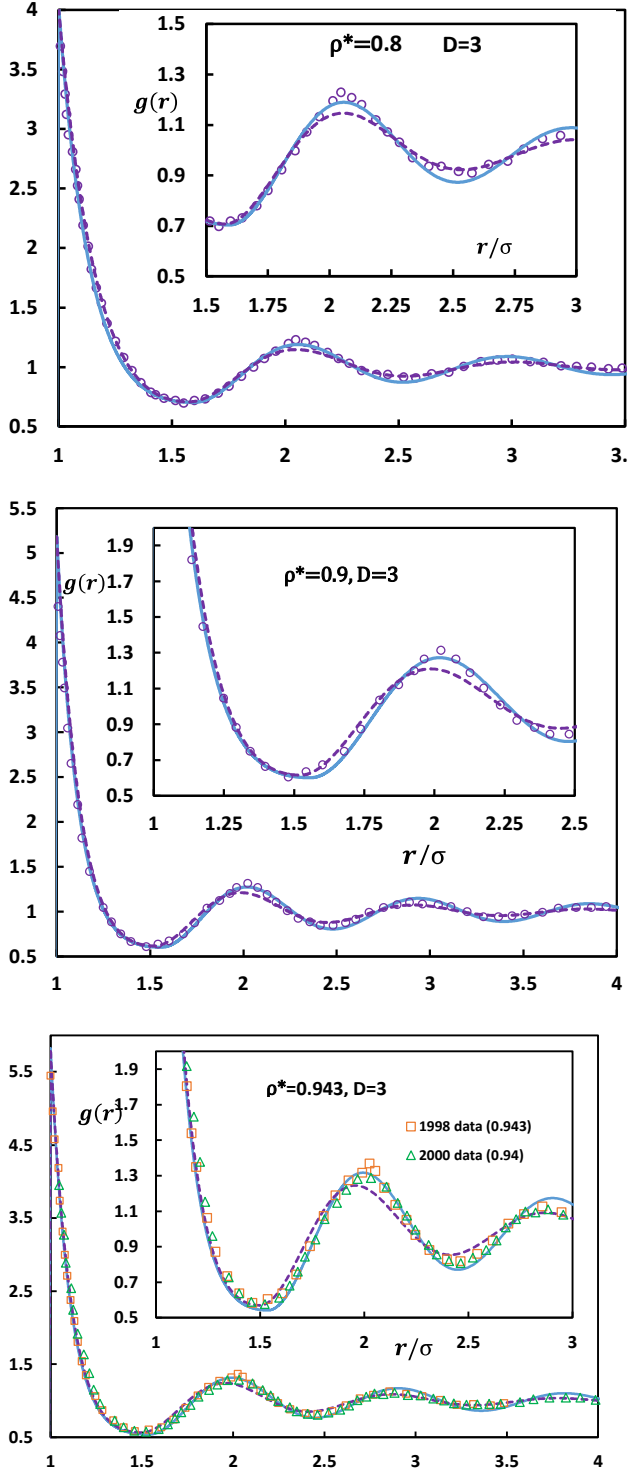

**Figure 8**. Comparisons of the 3D RDF calculated by the models (solid lines: this work, dashed lines: the TNJH RDF [10]) with the simulation data (points) at selected densities. Data sources are: $\rho^* = 0.3 - 0.9$ from ref.[10], and $\rho^* = 0.943, 0.94$ from ref.[20,21], respectively.

## V. A test of the 2D RDF with a classic perturbation theory

As mentioned, an important application of the RDF is in the perturbation theory [1,2,3]. Ref.[2] discussed discrete perturbation theory for the 2D square-well and the 2D Lennard-Jones (LJ) fluids using the 2D RDF proposed by Yuste and Santos [12]. For demonstration purpose, here we only consider a primitive perturbation theory for the 2D LJ fluid with the following potential:

$$u(r) = 4\epsilon \left[ \left(\frac{\sigma}{r}\right)^{12} - \left(\frac{\sigma}{r}\right)^6 \right] \quad (46)$$

where $\epsilon$ is the energy parameter (well-depth). With a classic perturbation theory the free energy of the LJ fluid is written as [2,3]

$$a = a^{id} + a_{ex}^{HD} + \beta a_1 + \beta^2 a_2 \quad (47)$$

where $\beta = 1/k_B T$. The ideal gas contribution to the free energy is given by:

$$\beta a^{id} = ln\rho^* + f(T) \quad (48)$$

where $f(T)$ is temperature-dependent only and can be ignored in a vapor-liquid equilibrium (VLE) calculation at a given temperature. As shown in Eq.(47) the contribution from the hard disk fluid, $a_{ex}^{HD}$, is taken as the reference. Considering the density range discussed here, we only take the virial contribution term in Eq.(27), $Z_v$, which yields:

$$\beta a_{ex}^{HD} = \int_0^\eta \frac{Z_v - 1}{\eta} d\eta = \frac{188\eta - 126\eta^2 - 13\eta^4}{52(1-\eta)^2}$$
$$- \frac{5}{13} ln(1-\eta) \quad (49)$$

The first perturbation term reads [1-3]:

$$\beta a_1 = \pi \rho \int_\sigma^\infty u(r) g^{HD}(r) r dr \quad (50)$$

and the second perturbation term is written as [2]:

$$\beta a_2 = -\frac{\pi}{2} \chi_T \rho \frac{\partial}{\partial \rho} \left[ \rho \int_\sigma^\infty u^2(r) g^{HD}(r) r dr \right] \quad (51)$$

Where $\chi_T = 1/(Z + \eta Z')$ is calculated with the EoS, Eq.(27):

$$Z' = \frac{dZ_v}{d\eta} = \frac{2 + \frac{1}{4}\eta + \frac{1}{6}\eta^2 - \frac{103}{126}\eta^3 + \frac{8}{21}\eta^4}{(1-\eta)^3} \quad (52)$$

Now we have an analytical expression for $g^{HD}(r)$, and the following integrations can be carried out numerically:

$$I_1 = \int_\sigma^\infty u(r) g^{HD}(r) r dr \quad (53)$$

$$I_2 = \int_\sigma^\infty u^2(r) g^{HD}(r) r dr \quad (54)$$

The numerical results are then fitted with polynomial functions in the density range, $0.4 < \rho^* \leq 0.85$:

$$I_1 = -0.576405 - 0.448057\rho^* - 0.068391\rho^{*2} + 0.20018$$

$$I_2 = 0.264217 + 0.536796\rho^* - 0.130209\rho^{*2} - 0.021598\rho$$

Therefore, the first and second perturbation terms can be written as, respectively:

$$\beta a_1 = \pi \rho I_1 \quad (57)$$

$$a_2 = -\frac{\pi}{2} \chi_T \rho \frac{\partial}{\partial \rho}(\rho I_2) \quad (58)$$

With Eq.(57) and (58), the total free energy, Eq.(47), and hence an EoS can be obtained finally. The compressibility and chemical potential ($\mu$) of the 2D LJ fluid are given by, respectively:

$$Z = \frac{P}{\rho k_B T} = \rho \frac{\partial \beta a}{\partial \rho}, \quad \beta \mu = \beta a + Z \quad (59)$$

At a given temperature, $T$, the VLE (calculations of saturated pressure and densities) of the LJ fluid can be determined by the equilibrium conditions:

$$P^v = P^l; \quad \mu^v = \mu^l \quad (60)$$

where the superscripts "v" and "l" refer to vapour and liquid phases, respectively. One can first calculate the saturated (equilibrium) pressure, $P^*$, in the $\mu - P$ plane, then determine the saturated vapor and liquid densities in the $P - \rho$ plane.

Figure 9 plots the results for the compressibility (pressure)-density relation at different temperatures (also known as the PVT relation) by using the EoS based on the simple perturbation theory, Eq.(47) and Eq.(59). The figure shows that the pressure predictions at $\rho^* = 0.5, 0.7, and\ 1.0$ reasonably agree with simulation results, while at high temperature, $T^* = 2.0$ the perturbation theory becomes less accurate for the high density range.

Figure 10 illustrates the saturated vapor pressure from the VLE calculation. Considering the simplicity of the theory, the agreement is satisfactory.

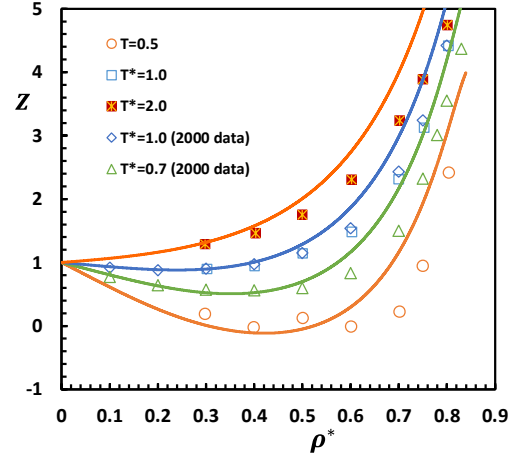

**Figure 9**. Compressibility (Pressure) -density (PVT) relation for the 2D LJ fluid. Data sources are Ref.[21] (2000 data) and Ref.[2] (the others).

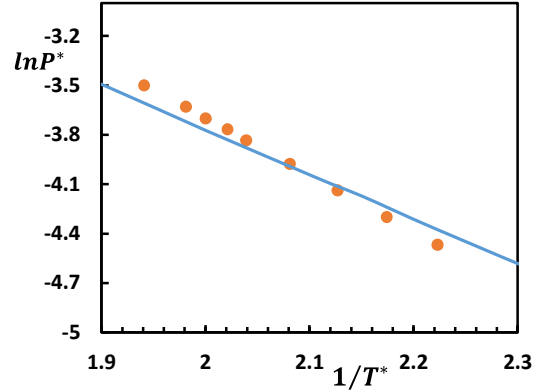

**Figure 10**. Saturated vapor pressure vs temperature for the 2D LJ fluid ($T^* = k_B T/\epsilon$). Solid line is the prediction of the perturbation theory. The simulation data are from Ref.[2].

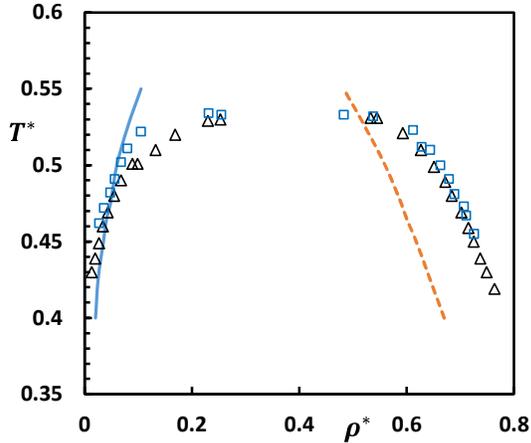

**Figure 11**. The VLE of the 2D LJ fluid in the $\rho^* - T^*$ plane. Lines are from prediction of the theory: the solid line is for the vapor phase and the dashed line for the liquid phase. The simulation data are from Ref.[2].

Finally, Figure 11 depicts saturated densities vs temperature from the VLE calculations. The agreement for the vapor density is acceptable, while for the liquid phase the deviations are significant. This is expected since it is a difficult task to predict the saturated liquid density with a simple EoS. For better description of the PVT relation and the VLE calculation, a more delicate theory is required, which is beyond the scope of this work.

## VI. Discussions and conclusions

Despite the importance of the radial distribution function of the hard disk fluid, a reliable expression over a wide density range is still missing until now. The primary goal of this work is to suggest an accurate RDF for the hard disk fluid over a wide density range up to the first-order phase transition point. As shown in sections III and IV, the general behavior of the RDFs are similar for both 2D and 3D fluids, which allows us to use a similar expression for both cases. We propose a simple expression based on the model suggested by Trokhymchuk et al. for the hard sphere fluid [10]. Compared with the TNJH expression, the new RDF has less parameters and the final expressions are remarkably accurate for both 2D and 3D fluids when the accurate equations of state [13,14] are used as inputs. As a result of the simplification of the depletion branch, the Lambert W function, Eq.(20), helps reducing an unknown in the global optimization process.

The new 2D RDF successfully reproduces the structural properties over a wide density range, and even indicates a long-range order as the system approaches to the hexatic phase. Therefore, the 2D RDF is suggested for the entire density range: $\rho^* = 0.4 - 0.9$. To the best of the author's knowledge, this is only 2D RDF reported so far for the hard disk fluid over such a wide density range. This 2D RDF lays a foundation for future developments of more accurate perturbation theories [9] and analytical RDF for the 2D LJ fluid.

For the 3D RDF, the new expression is recommended for the density up to the freezing point, $\rho^* = 0.943$. Even with considerable simplifications, the final RDF is more accurate than the original TNJH expression [10] in high density range. The simple math form makes it a good candidate for various applications, such as for deriving a simple and analytical RDF for the LJ fluid and for perturbation theories.

With the 2D LJ fluid as an example, the new 2D RDF has been applied to the predictions of the thermodynamic properties by using a primitive perturbation theory and the results are promising. As expected, the classic perturbation theory, Eq.(47), exhibits limited capacity in predictions of the PVT and VLE behaviors. Consequently, some sophisticated theories are required for highly accurate predictions.

## Appendix A: 2D fluid: constants of Eq.(27) and phase transition predicted by the EoS

For more details on the EoS and phase transitions, the reader is referred to Refs. [13,14].

**Table A1** EoS constants in Eq.(27)

| $b_1$ | $b_2$ | $m_1$ | $m_2$ | $1/c$ |
|---|---|---|---|---|
| $-1.04191 \times 10^8$ | $2.66813 \times 10^8$ | 53 | 56 | 0.75 |

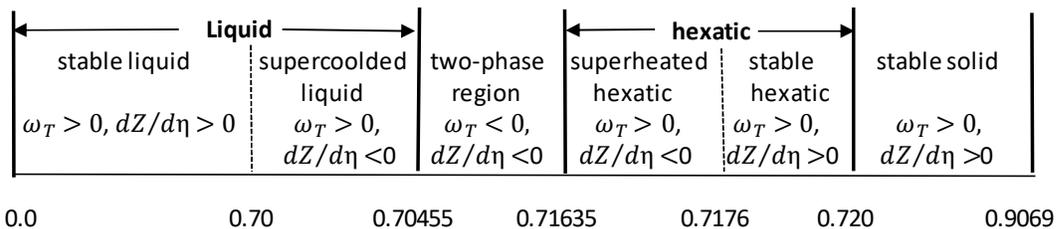

| | Liquid | | | hexatic | | |
|---|---|---|---|---|---|---|
| stable liquid | supercoolded liquid | two-phase region | superheated hexatic | stable hexatic | stable solid |
| $\omega_T > 0, dZ/d\eta > 0$ | $\omega_T > 0, dZ/d\eta < 0$ | $\omega_T < 0, dZ/d\eta < 0$ | $\omega_T > 0, dZ/d\eta < 0$ | $\omega_T > 0, dZ/d\eta > 0$ | $\omega_T > 0, dZ/d\eta > 0$ |
| 0.0 | 0.70 | 0.70455 | 0.71635 | 0.7176 | 0.720 | 0.9069 |

**Figure A1.** Summary of all phases over the entire density range [14]. The numbers at the bottom are the packing fractions, η.

In Figure A1, the rigidity is defined as the work required to increase the density, $\omega_T = (dP/d\rho)_T = (d\bar{G}/d\rho)_T$, where $\bar{G}$ is the Gibbs free energy. According to the fluctuation theory, rigidity is inversely proportional to an average dimensionless variance in total number of particles (*N*) at constant volume and is given by the following:

$$\omega_T = \frac{N_a k_B T}{[\langle(\Delta N)^2\rangle]_{V,T}} = Z + \eta \frac{dZ}{d\eta} = \chi_T^{-1} \quad (A1)$$

Namely, the rigidity is the inverse of the reduced isothermal compressibility.

## Appendix B: Recipes to calculate RDF for hard disk and hard sphere fluids

**Generic expressions:**

$$g(r) = \begin{cases} 0, & r < \sigma \\ \frac{A}{r}e^{\mu(r-\sigma)} + \frac{B}{r}e^{\alpha(r-\sigma)}, & \sigma \leq r < r_m \\ 1 + \frac{C}{r}\cos(\omega r + \delta)e^{-\kappa r}, & r_m \leq r < \infty \end{cases} \quad (B1)$$

$$\frac{B}{\sigma} = \frac{r^* g_m - g(\sigma)e^{\mu\sigma(r^*-1)}}{e^{\alpha\sigma(r^*-1)} - e^{\mu\sigma(r^*-1)}} \quad (B2)$$

$$\frac{A}{\sigma} = g(\sigma) - \frac{B}{\sigma} \quad (B3)$$

$$\delta = -\omega\sigma r^* - \arctan\frac{\kappa\sigma r^* + 1}{\omega\sigma r^*} \quad (B4)$$

$$\frac{C}{\sigma} = \frac{r^*[g_m - 1]e^{\kappa\sigma r^*}}{\cos(\omega\sigma r^* + \delta)} \quad (B5)$$

where $g(\sigma) = (Z-1)/2^{D-1}\eta$, from an EoS.

**Parameters for the 2D RDF:**

$$\omega\sigma = 4.07688 + 3.53995\eta \quad (B6)$$

$$\kappa\sigma = 1.19878 - 1.54059\eta - 0.567808\eta^2 \quad (B7)$$

$$\alpha\sigma = -1.69377 + 13.50042\eta - 29.8772\eta^2 \quad (B8)$$

$$r^* = 2.14098 - 1.03267\eta + 0.320164\eta^2 + 0.0102421\eta^3 \quad (B9)$$

$$g_m = 1.97652 - 6.34821\eta + 11.53962\eta^2 - 5.96960\eta^3 - 3.92125\eta^4 \quad (B10)$$

$$\mu\sigma = -116.2837 + 1218.065\eta - 4784.826\eta^2 \\ +9229.068\eta^3 - 8862.922\eta^4 + 3422.542\eta^5 \quad (B11)$$

where $\eta = \pi\rho^*/4$ and the RDF at contact for the 2D fluid reads:

$$g(\sigma) = \frac{1 - \frac{7}{16}\eta + \frac{1}{36}\eta^2 - \frac{2}{21}\eta^3}{(1-\eta)^2} + \frac{b_1\eta^{m_1} + b_2\eta^{m_2}}{1 - c\eta} \quad (B12)$$

**Parameters for the 3D RDF:**

$$\omega\sigma = 6.56745 + 0.721795\eta \quad (B13)$$

$$\kappa\sigma = 1.75879 - 3.23977\eta + 0.449588\eta^2 - 0.148890\eta^3 \quad (B14)$$

$$\alpha\sigma = 0.595157 - 2.469402\eta - 31.99603\eta^2 - 0.897749\eta^3 \quad (B15)$$

$$r^* = 1.960492 - 0.240131\eta - 3.26102\eta^2 + 4.08014\eta^3 \quad (B16)$$

$$g_m = 1.072954 - 1.13263\eta + 3.05087\eta^2 - 5.94853\eta^3 \quad (B17)$$

$$\mu\sigma = 0.58236 + 8.43296\eta - 34.52268\eta^2 + 58.63440\eta^3 \\ -35.13525\eta^4 \quad (B18)$$

where $\eta = \pi\rho^*/6$ and the RDF at contact for the 3D fluid reads:

$$g(\sigma) = \frac{1 - \frac{1}{2}\eta + \frac{5}{52}\eta - \frac{1}{4}\eta^3 + \frac{1}{8}\eta^4}{(1-\eta)^3} \quad (B19)$$